\documentclass[prd,aps,preprint,tightenlines,superscriptaddress]{revtex4}
\usepackage{amsfonts}
\usepackage{mathrsfs}
\usepackage{amssymb}
\usepackage{epsfig}
\usepackage{amsmath}
\begin{document}
\title{Constraining Right-Handed Scale Through Kaon Mixing \\in Supersymmetric Left-Right Model}
\author{Yue Zhang}
\affiliation{Center for High-Energy Physics and Institute of
Theoretical Physics, Peking University, Beijing 100871, China}
 \affiliation{Maryland Center for Fundamental Physics, Department of Physics, University of
Maryland, College Park, Maryland 20742, USA }
\author{Haipeng An}
 \affiliation{Maryland Center for Fundamental Physics, Department of Physics, University of
Maryland, College Park, Maryland 20742, USA }
\author{Xiangdong Ji}
 \affiliation{Maryland Center for Fundamental Physics, Department of Physics, University of
Maryland, College Park, Maryland 20742, USA } \affiliation{Center
for High-Energy Physics and Institute of Theoretical Physics, Peking
University, Beijing 100871, China}
\date{\today}
\begin{abstract}
We study flavor-changing neutral current and CP violations in the minimal supersymmetric
left-right model. We calculate the beyond-standard-model contributions to the neutral
kaon mixing $\Delta M_K$ and $\epsilon$, and find possible to have a numerical cancelation
between the contributions from the right-handed gauge boson and supersymmetric box
diagram. With the cancelation, the right-handed $W$-boson mass
scale can be lowered to about 2 TeV, well within the search limit of LHC.

\end{abstract}
\maketitle

Physics beyond the standard model (SM) has been actively pursued by both theorists and
experimentalists for decades. New physics is needed at TeV scale to explain a number of
salient features of SM, particulary the origin of the scale at which the electroweak
symmetry is spontaneously broken. From the phenomenological point of view, candidates of
new physics have to confront constraints from the existing low-energy data. Flavor-changing 
neutral currents (FCNC) and the combined charge-conjugation and parity (CP)
violation impose some of the most stringent constraints. In the SM, FCNC is highly
suppressed due to the celebrated Glashow-Iliopoulos-Maiani (GIM) mechanism and the fact
that the Cabibbo-Kobayashi-Maskawa (CKM) matrix is nearly diagonal. CP violation in the light-quark 
sector is small because the Dirac phase appears only in off-diagonal elements of
the CKM matrix. In sundry extensions of SM, there are generically new sources of FCNC and
CP violations. It is necessary to investigate their implications on phenomena such as
neutral kaon mixing and neutron electric dipole moment (EDM), before testing them at
colliders.

The left-right symmetric model (LRSM) \cite{lrmodel} was introduced to restore parity
symmetry at high-energy: Parity is considered as a good symmetry when the energy scale is
sufficiently high and is broken spontaneously at low energy. Besides the fundamental
symmetry \cite{lee}, one of the nice features of the model is the electric charge
quantization: the obscured hypercharge in SM is explained in terms of the baryon and
lepton numbers and the left-right (LR) isospins. In the minimal LRSM, the
right-handed quarks are doublets under the $SU(2)_R$ gauge group, so the mixing among the
right-handed quarks becomes physical observables. There is a corresponding right-handed CKM matrix
$V^R_{\rm CKM}$, analogous to the SM CKM matrix $V^L_{\rm CKM}$. Physics of CP violation
in LRSM is quite interesting. The strong CP problem is solved by eliminating the
dimension-four gluon operator $G\tilde G$ by parity at high-energy. 
In the weak sector, two special scenarios have been generally discussed historically. 
One is called ``manifest" LR
symmetry without spontaneous CP violation, in which parity guarantees left-handed and
right-handed CKM matrices are identical $V^R_{\rm CKM}=V^L_{\rm CKM}$. The other is called
``pseudo-manifest" LR symmetry~\cite{scpv} such that the Lagrangian is invariant
under P and CP, both of which are broken spontaneously. $V^R_{\rm CKM}$ is then proportional
to the complex conjugate of the $V^L_{\rm CKM}$ multiplied by additional CP phases, determined by
those of the higgs vacuum expectation values (vev). As was pointed out by the
authors in a recent paper \cite{our paper}, neither of the above scenarios is realistic.
In that work, the explicit form of $V^R_{\rm CKM}$ has been solved in the presence of
both CP violating Lagrangian and higgs vev's, and is related to the $V^L_{\rm CKM}$ and the spontaneous CP
phases in a non-trivial manner.

One of the important issues in the LR model has been the scale at which the
right-handed current interaction becomes significant. Constraints on the righthanded
W-boson mass $M_{W_R}$ in the non-supersymmetric version of LRSM
from CP asymmetries have been explored in Ref. \cite{our paper}.
The most well-known and stringent lower bound on $M_{W_R}$ comes from the mass difference
between $K_L$ and $K_S$, in which the $W_L-W_R$ box-diagram contribution is enhanced by
both the Wilson coefficient and hadronic matrix element, altogether by a factor of ${\cal
O}(10^3)$. A re-evaluation of this contribution with updated values of strange quark mass
and hadron matrix element and consideration of CP violating observables yield
a lower bound $M_{W_R} > 4.0$ TeV. The current
experimental bound on $M_{W_R}$ in direct collider search is about 800 GeV \cite{PDG}.
Given these, is it possible to have a right-handed gauge-boson with mass of order 1-2 TeV 
which can be copiously produced at LHC while satisfying various low-energy constraints?

In this paper, we explore the above question in the context of the supersymmetric
(SUSY) LRSM, where there is an additional contribution to the kaon mixing from the
sparticle box-diagrams. As we shall see, the numerical cancelation among the new
contributions can happen, and as a consequence, the low-energy constraint on the
left-right symmetric scale can be relaxed. A previous analysis of the kaon mixing in
the same model has been carried out in \cite{Frank:2003yi}, deriving the constraints on
the squark flavor-mixing parameters. Here we focus on a different issue---the interplay 
between the $W_L-W_R$ mixing and SUSY box-diagram contributions---finding a minimal bound on 
the right-handed scale. We note that the same question has been explored in the context 
of two Higgs-bidoublet model without SUSY \cite{Wu:2007kt}, and the answer depends 
crucially on the size of the charged Higgs particle mass which shall have the scale $M_{W_R}$
if SM is to be recovered in $M_{W_R}\rightarrow \infty$ limit. In our case, the charged Higgs
contribution through the box-diagram is small \cite{Basecq:1985cr}, whereas the tree-level
neutral higgs (FCNH) contribution can be large and will be considered.

This paper is organized as follows. After a brief introduction to the SUSY LRSM, we review the
argument that CP violation in the model must be manifest, and as a by-product, the neutron EDM is naturally
suppressed. However, $\epsilon_K$ then imposes a very stringent bound from
the $W_L-W_R$ box-diagrams, $M_{W_R}>15$ TeV in the non-SUSY model. Next, we include the
contribution from the gluino box-diagrams assuming the FCNH's are very heavy and a minimal supergravity (mSUGRA) 
boundary condition. We present the numerical evidence for the cancelation among the
new contributions so that the right-handed mass scale can be lowered to the 2 TeV range.
The cancelation can be complete if one assumes generic squark mass-insertions,
even when the FCNH is light. In this case, there is no theory bound on the right-handed scale 
from neutral kaon system. 

The minimal SUSY LRSM is based on the gauge group $SU(2)_L \times
SU(2)_R \times U(1)_{B-L} \times P$. Right-handed neutrinos are introduced for each
family, and all right-handed fermions form doublets under $SU(2)_R$ gauge group:
\begin{eqnarray}
Q_L &=& \left( \begin{array}{c} u_L \\ d_L \\ \end{array} \right) \in \left(2, 1,
\frac{1}{3}\right), \ Q_R^c = \left(
\begin{array}{c} u_R^c
\\ d_R^c \\ \end{array} \right) \in \left(1, 2,
-\frac{1}{3}\right) \nonumber\\
L_L &=& \left( \begin{array}{c} \nu_L \\
l_L \\ \end{array} \right) \in \left(2, 1, -1\right), \ L_R^c = \left(
\begin{array}{c} \nu_R^c
\\ l_R^c \\ \end{array} \right) \in \left(1, 2, 1\right) \ ,
\end{eqnarray}
where the gauge quantum numbers $SU(2)_L \times SU(2)_R \times U(1)_{B-L}$ are listed in
the brackets and $Q_R^c$ lies in the conjugate representation of $Q_R$ etc. The
renormalizable superpotential is
\begin{eqnarray}\label{super}
W &=& Y_a ~Q_L^T \tau_2 \Phi_a \tau_2 Q_R^c + Y_a^l~ L_L^T \tau_2 \Phi_a
\tau_2 L_R^c \nonumber \\
&+& f L_L^T i \tau_2 \Delta_L L_L + f_c L_R^{cT} i \tau_2 \Delta_R^c L_R \nonumber \\
&+& \frac{1}{2} \mu_{ab}{~\rm Tr}\left( \tau_2 \Phi_a^T \tau_2 \Phi_b \right)
+\mu_L{~\rm Tr}\left( \Delta_L \delta_L \right) \nonumber \\
&+& \mu_R{~\rm Tr}\left( \Delta_R^c \delta_R^c \right)\ ,
 \end{eqnarray}
where $a,b=u,d$. In the minimal model, two distinct Higgs bidoublets, $\Phi_u$ and
$\Phi_d$, and associated Yukawa couplings are needed to generate a nontrivial CKM mixing due to holomorphism of $W$.
The triplet Higgs fields are introduced to give neutrino Majorana masses.
They have to be doubled to guarantee $U(1)_{B-L}$ anomaly free.
The Higgs bidoublets $\Phi_{u,d}$ and triplets $\Delta_{L,R}$, $\delta_{L,R}$ in
(\ref{super}) are defined as
\begin{eqnarray}
\Phi_{a} &=& \left( \begin{array}{cc} \Phi_{a1}^0 & \Phi_{a2}^+ \ ,  \\
\Phi_{a1}^- & \Phi_{a2}^0 \end{array} \right) \in \left(2, 2, 0\right)
\ , \nonumber\\
\Delta_L &=& \left( \begin{array}{cc} \Delta_{L}^+ / \sqrt{2} & \Delta_{L}^{++} \\ \Delta_{L}^{0} & - \Delta_{L}^+ / \sqrt{2}\\
\end{array} \right) \in \left(3, 1, 2\right) \ , \nonumber \\
\delta_L &=& \left( \begin{array}{cc} \delta_{L}^- / \sqrt{2} & \delta_{L}^{0} \\ \delta_{L}^{--} & - \delta_{L}^- / \sqrt{2}\\
\end{array} \right) \in \left(3, 1, -2\right) \ , \nonumber \\
\Delta_R^c &=& \left( \begin{array}{cc} \Delta_{R}^- / \sqrt{2} & \Delta_{R}^{--} \\ \Delta_{R}^{0} & -\Delta_{R}^- / \sqrt{2} \\
\end{array} \right) \in \left(1, 3, -2\right) \ , \nonumber\\
\delta_R^c &=& \left( \begin{array}{cc} \delta_{R}^+ / \sqrt{2} & \delta_{R}^{0} \\ \delta_{R}^{++} & - \delta_{R}^+ / \sqrt{2}\\
\end{array} \right) \in \left(1, 3, 2\right) \ .
\end{eqnarray}

The soft SUSY breaking terms involving the scalar (spinor) components of chiral
(vector) superfields read
\begin{eqnarray}
\mathscr{V}_{\rm soft} &=& \widetilde q_L^\dag m_{\widetilde Q_L}^2 \widetilde q_L +
\widetilde q_R^{c\dag} m_{\widetilde Q_R}^2
\widetilde q_R^c \nonumber \\
& + & \widetilde l_L^\dag m_{\widetilde l_L}^2 \widetilde
l_L + \widetilde{l_R}^{c \dag} m_{\widetilde l_L}^2 \widetilde{l_R}^c + m^2_{\phi_{ab}}{~\rm Tr}(\phi_a^\dag \phi_b) \nonumber \\
& + & m_{\Delta_L}^2 {~\rm Tr}(\Delta_L^\dag \Delta_L) +
m_{\delta_L}^2{~\rm Tr}(\delta_L^\dag \delta_L) \nonumber \\
& + & m_{\Delta_R}^2{~\rm Tr}(\Delta_R^{c\dag} \Delta_R^c) +
m_{\delta_R}^2{~\rm Tr}(\delta_R^{c\dag} \delta_R^c) \nonumber \\
& + & \left[ \frac{1}{2} \left(  M_L \lambda_L^a \lambda^a_L + M_R
\lambda_R^a \lambda^a_R + M_V \lambda_V \lambda_V + M_3 \lambda_g
\lambda_g \right) \right.  \nonumber\\
& + & \left. \widetilde q_L^T \tau_2 A_u \phi_u \tau_2 \widetilde
q_R^c + \widetilde q_L^T \tau_2 A_d \phi_d \tau_2 \widetilde q_R^c +
\widetilde l_L^T \tau_2 A_l \phi \tau_2 \widetilde l_R^c \right. \nonumber \\
& + & \left. i v \left( \widetilde l_L^T \tau_2 \Delta_L \widetilde l_L + \widetilde
l_R^{cT} \tau_2 \Delta_R^c \widetilde l_R^c \right) + h.c. \right] \nonumber \\
& + & \left[  B \mu_{ab} {\rm Tr}\left( \tau_2 \phi_a^T \tau_2 \phi_b
\right) + B_L \mu_L {\rm Tr}\left( \Delta_L \delta_L \right) \right. \nonumber \\
& + & \left.  B_R \mu_R {\rm Tr}\left( \Delta_R^c \delta_R^c \right)  + {\rm h.c.}
\right]
\end{eqnarray}
Under LR symmetry, one can define the transformation of the superfields \cite{susylr}
\begin{eqnarray}
&&Q_L \leftrightarrow Q_R^{c*}, \ \Phi_{u,d} \leftrightarrow
\Phi_{u,d}^{\dag},\ \Delta_L \leftrightarrow \Delta_R^{c*} \nonumber \\
&& W_L \leftrightarrow W_R^{*}, \ \theta \rightarrow \overline
\theta
\end{eqnarray}
where $W_{L, R}$ are the gauge vector superfields.  The lagrangian is invariant under
LR symmetry implies that
\begin{eqnarray}\label{hermitian}
h_{u, d} &=& h_{u, d}^{\dag}, \ \mu_{ab} = \mu_{ab}^*, \ \mu_L =
\mu_R^*, \ M_L = M_R^*, \nonumber \\
\ M_V &=& M_V^*, \ M_3 = M_3^*, \ A_u = A_u^{\dag}, \ A_d =
A_d^{\dag}, \nonumber \\
m_{\widetilde Q_L}^2 &=& m_{\widetilde Q_R}^{2\dag}, \
m_{\Delta_L}^2 = m_{\Delta_R}^2, \ m_{\delta_L}^2 = m_{\delta_R}^2, \nonumber \\
m_{\phi_{ab}}^2 &=& m_{\phi_{ba}}^2 \in \mathbb{R}, \ B = B^*, \ B_{L} = B_R^*, \ g_L=
g_R \ ,
\end{eqnarray}
where $g_L$ and $g_R$ are the gauge coupling constants. The parity symmetry eliminates
most of the phases in the soft parameters.

The gauge symmetry is spontaneously broken at a right-handed scale $v_R$ to the SM gauge group.
We now briefly review the status of spontaneous breaking of $SU(2)_R \times U(1)_{B-L}$ down to $U(1)_Y$. It is well known that in the minimal model, spontaneous parity breaking does not
happen \cite{susylr2}. Depending on the scale at which LR symmetry is restored, one can augment the minimal model in several
different ways to obtain the symmetry breaking. If the right-handed scale is high, one can either include non-renormalizable terms in the superpotential or extend the Higgs sector, introducing additional B-L neutral triplets \cite{Aulakh:1997ba}. The additional couplings will pick up the desired Higgs condensate in Eq. (\ref{define}) as the true vacuum. The most economic and only possible way for
low scale LR symmetry is to introduce R-parity violation, giving right-handed sneutrino a condensate $\langle \widetilde \nu_R^c \rangle$. In this scenario, there is an upper bound on the right-handed scale and a corresponding lower bound on the sneutrino condensate \cite{susylr2, RpVpaper}
\begin{eqnarray}
v_R \leq \frac{v}{2f}, \ \ \ \langle \nu^c_R \rangle \geq \frac{v_R}{4 \pi} \left[ \log \frac{M_{\rm gut}}{m_{\rm SUSY}} \right]^{\frac{1}{2}} \ ,
\end{eqnarray}
where $f$ and $v$ are couplings in Eqs. (2) and (4), $M_{\rm gut}$ is a high scale beyond which the SUSY LRSM unifies into a larger theory,
and $m_{\rm SUSY}$ is the SUSY breaking scale.
The right-handed sneutrino condensation brings about R-parity violation in the bilinear form $W_{\rm RpV} \propto L_i H_d$. This will lead to many interesting phenomenology in cosmology and at colliders. We will address these issues elsewhere.

After spontaneous LR and electroweak symmetry breaking, the Higgs vev's take the following general forms
\begin{eqnarray}\label{define}
\langle \phi_u \rangle &=& \left( \begin{array}{cc} \kappa_u' & 0\\
0 & \kappa_u  \end{array} \right), \  \langle \phi_d \rangle = \left( \begin{array}{cc} \kappa_d  & 0 \\
0 & \kappa_d' \end{array} \right) \nonumber \\
\langle \Delta_R^c \rangle &=& \left( \begin{array}{cc} 0 & 0 \\
v_R &  0 \end{array} \right), \
\langle \delta_R^c \rangle = \left( \begin{array}{cc} 0 & \bar v_R \\
0 &  0 \end{array} \right) \nonumber \\
\langle \Delta_L \rangle &=& \langle \delta_L \rangle =0 \ ,
\end{eqnarray}
At tree level, we have $v_R \approx \bar v_R$ due to the D-flat condition, and we will assume $\langle H_{d1} \rangle = \kappa_u' = 0$ and $\langle H_{u2} \rangle = \kappa_d' = 0$ for simplicity.
It has first been proved that vev of the higgs bidoublets must be real in the SUSY SM with four Higgs doublets in Ref. \cite{Masip:1995sm}, and later extended to minimal SUSY LRSM in Ref. \cite{susylr}.
A remarkable feature of the SUSY LRSM is that it provides a natural
solution to both the SUSY and strong CP problems. The relevant phases for CP violation at low energy are the combinations ${\rm arg} (A m_{1/2}^*)$ and
${\rm arg} (B m_{1/2}^*)$ \cite{Dugan:1984qf}. As observed in \cite{susylr, Babu:2001se}, due to parity invariance, they vanish up to one-loop level, provided $M_L =
M_R$ are real at the SUSY breaking scale. In fact, one can further extend the LRSM to $SO(10)$ grand unified theories (GUT), where $M_L$ and $M_R$ unify into real universal gaugino mass. The finite contributions only appear contribute at least at two-loop order. It has been established that the contribution to $\overline \Theta$ is at most on the order $10^{-8}-10^{-10}$ \cite{Babu:2001se}. This calls for much milder fine-tuning of parameters compared to the unconstrained version of minimal supersymmetric standard model (MSSM).

Corresponding to Eq. (\ref{define}), $\langle H_{u1} \rangle = \kappa_u$, $\langle H_{d2} \rangle = \kappa_d$. The quark mass matrices are
\begin{eqnarray}\label{mass}
M_u = Y_u^* \kappa_u, \ \ \ M_d = Y^*_d \kappa_d
\end{eqnarray}
and $\tan\beta = \displaystyle \frac{\kappa_u}{\kappa_d}$. From the hermiticity of the Yukawa couplings and the fact that $\kappa_u$, $\kappa_d$ are real numbers, one concludes CP violation in the quark sector is quasi-manifest~\cite{Kiers:2002cz}
\begin{eqnarray}\label{10}
V_{Rij}^{\rm CKM} = \pm V_{Lij}^{\rm CKM} \ ,
\end{eqnarray}
where the sign provides the opportunity for cancelation. Again, there is no spontaneous CP violation phase.

In the remainder of this paper, we study the low-energy flavor mixing and CP-violation phenomenology
in the neutral kaon system in the SUSY LRSM. We focus on the mass difference between $K_L$ and
$K_R$ and the indirect CP violation parameter $\epsilon_K$. Two significant beyond-SM contributions
to the $K^0-\overline K^0$ mixing are the $W_L-W_R$ box-diagram and the gluino exchange diagrams.
Our aim is to explore to what extend the right-handed scale $v_R$ can be lowered.

The $K_L-K_S$ mass difference and indirect CP violation are calculated using the standard
formula
\begin{eqnarray}
\Delta m_K &=& 2 {~\rm Re} \langle K^0 | \mathscr{H}^{\Delta S=2}_{\rm eff}
| \bar{K}^0 \rangle \ , \nonumber \\
\epsilon_K &=&  \frac{e^{i \pi/4}}{\sqrt{2}} \frac{~{\rm Im} \langle K^0 |
\mathscr{H}^{\Delta S=2}_{\rm eff} | \bar{K}^0 \rangle}{ \Delta m_K } \ ,
\end{eqnarray}
where $\mathscr{H}^{\Delta S=2}_{\rm eff} \equiv\mathscr H_{12}^{SM} +\mathscr H_{12}^{LR} +\mathscr
H_{12}^{\widetilde g}$ with $\mathscr H_{12}^{SM}$ is the SM contribution, and $\mathscr H_{12}^{LR}$,
$\mathscr H_{12}^{\widetilde g}$ and other neglected contributions will be explained in below.

In the LRSM, due to the absence of GIM
suppression the $W_L-W_R$ box-diagram makes important contribution to kaon mixing.
\begin{eqnarray}\label{c1}
\mathscr{H}^{LR}_{12}&=&\frac{G_F}{\sqrt{2}}\frac{\alpha}{4\pi\sin^2\theta_W}2\eta\lambda^{LR}_i\lambda^{RL}_j\sqrt{x_ix_j}\left[(4+\eta
x_ix_j) \right. \nonumber\\
& &\times \left. I_1(x_i,x_j,\eta) - (1+\eta)I_2(x_i,x_j,\eta)\right] f^2_K \eta_4 B_4 \left( \frac{m_K}{m_s+m_d} \right)^2  + {\rm h.c.} \ ,
\end{eqnarray}
in which $\lambda^{LR}_{i}=V^{CKM *}_{Li2}V^{CKM}_{Ri1}$,
$f_K=130$ MeV, the QCD leading-logarithmic evolution factor $\eta_4 = 1.4$ and
$B$-factor for the four-quark matrix element $B_4 =0.81$ at energy scale $\nu = 2$ TeV,
$\displaystyle\eta=\left(\frac{M_L}{M_R}\right)^2$,
$\displaystyle x_i=\left(\frac{m_i}{M_L}\right)^2$, $i=u,c,t$ and
\begin{eqnarray}\label{c2}
I_1(x_i,x_j,\eta)&=&\frac{\eta\ln(1/\eta)}{(1-\eta)(1-x_i\eta)(1-x_j\eta)} + \left[\frac{x_i\ln
x_i}{(x_i-x_j)(1-x_i)(1-x_i\eta)}+(i\leftrightarrow j)\right] \ ,
\nonumber\\
I_2(x_i,x_j,\eta)&=&\frac{\ln(1/\eta)}{(1-\eta)(1-x_i\eta)(1-x_j\eta)}+ \left[\frac{x_i^2\ln
x_i}{(x_i-x_j)(1-x_i)(1-x_i\eta)}+(i\leftrightarrow j)\right] \ .
\end{eqnarray}
The charm-quark exchange in the $W_L-W_R$ box-diagram makes dominant contribution to $K_L$-$K_S$ mass difference, yielding a lower bound of 3.5 TeV for $M_{W_R}$. In the SUSY LRSM with only manifest CP violation, the Dirac phase $\delta_{CP}$ is the sole source for $\epsilon_{K}$, and the $c-t$ quarks exchange in the box provides the
dominant contribution. If this were the only one, the experimental data on $\epsilon_K$ imposes a very stringent bound on the right-handed scale
\begin{eqnarray}
M_{W_R} > 15\ {\rm TeV} \ ,
\end{eqnarray}
The bound is much more stringent than that in the presence of spontaneous CP phase in non-SUSY LRSM. 

On the other hand, the dominant SUSY contribution to the neutral kaon mixing comes from the gluino
exchange diagrams. The relevant physics is similar to that of MSSM. The gluino-exchange box-diagram (as well as the crossed ones) gives
\begin{eqnarray}\label{gluinoH}
H_{12}^{\widetilde g} &=& -\ \frac{\alpha_s^2}{216 m_{\tilde{q}}^2} \frac{2}{3} m_{K}
f^2_{K} \times \nonumber \\
&+& \left\{ (\delta^d_{LL})^2_{12} \left[ \left( 384
\left( \frac{m_{K}}{m_d + m_s}\right)^2 + 120 \right) x f_6(x)
+ \left( -24 \left( \frac{m_{K}}{m_d + m_s}\right)^2 + 168
\right) \tilde{f}_6(x) \right] \right. \nonumber \\
&-& \left. (\delta^d_{LR})^2_{12} \left[ 264 \left( \frac{m_{K}}{m_d +
m_s}\right)^2 x f_6(x) + \left(  144
\left( \frac{m_{K}}{m_d + m_s}\right)^2 + 84 \right) \tilde{f}_6(x) \right]
\right\}\ ,
\end{eqnarray}
where $x$ is defined as squared ratio of gluino mass to averaged squark mass
$x=M_3^2/m_{\tilde{q}}^2$. We have simplified the corresponding formula in \cite{gluino} with the relations among mass insertions $(\delta^d_{LL})_{12}=(\delta^d_{RR})_{12}$ and $(\delta^d_{LR})_{12}=(\delta^d_{RL})_{12}$ under left-right symmetry. The functions $f_6(x)$, $\tilde{f}_6(x)$ are
\begin{eqnarray}
f_6(x) &=& \frac{6(1+3x)\ln x + x^3 - 9 x^2 - 9 x + 17 }{6(x-1)^5} \nonumber \\
\tilde{f}_6(x) &=& \frac{6x(1+x)\ln x - x^3 - 9 x^2 + 9 x + 1 }{3(x-1)^5} \ ,
\end{eqnarray}
In the super-KM basis to be described below, flavor violations are attributed to the off-diagonal elements of the squark-mass-squared matrices $\left( M^{~q~2}_{AB} \right)_{i\neq j}$.
As a concrete model for these parameters, we consider the mSUGRA model for SUSY breaking, 
where the flavor universality of squark masses is assumed at some high scale $M^*$. 
The universal squark and higgs boson masses are denoted by $m_0$, gaugino mass by $m_{1/2}$, and a universal $\mu$ parameter. The trilinear scalar couplings are assumed to be proportional to the corresponding Yukawa couplings $A_{u,d} = A_0 Y_{u,d}$. Then the flavor-violation in the squark masses is generated via renormalization group evolution effects starting from $M^*$.
The mass insertions are defined as $\left(\delta^q_{AB}\right)_{ij} \equiv
\left(M^{~q~2}_{AB}\right)_{ij} / m_{\widetilde q}^2\ (A,B=L,R)$. $m_{\tilde{q}}^2 \equiv {~\rm Tr} \left( M_{uLL}^2 + M_{uRR}^2 + M_{dLL}^2 + M_{dRR}^2 \right)/12$ is averaged squark mass squared. In MSSM, phenomenological bounds on the SUSY parameter spaces can be found in Ref. \cite{gluino, mssm}. When the multi-TeV right-handed scale is introduced in LRSM, we have to re-evaluate these bounds.

The renormalization group equations (RGE) for the soft masses and $A$-terms in the SUSY LRSM are known \cite{rge}
\begin{eqnarray}\label{rge}
16 \pi^2 \frac{d}{d t} m_{\widetilde Q_L}^2 &=& 2 m_{\widetilde Q_L}^2 Y_a Y_a^{\dag} +
Y_a \left[ 2 Y_a^\dag  m_{\widetilde Q_L}^2 + 4 Y_b^\dag m_{\phi a b}^2 + 4 m_{\widetilde
Q_R}^2 Y_a^\dag \right] \nonumber \\
&& + 4 A_{a} A_{a}^\dag - \frac{1}{3} M_V^2 g_V^2 - 6 M_L^2
g_L^2 - \frac{32}{3} M_3^2 g_3^2 + \frac{1}{8} g_V^2 S_3 \nonumber \\
16 \pi^2 \frac{d}{d t} m_{\widetilde Q_R}^2 &=& 2 m_{\widetilde Q_R}^2 Y_a^{\dag} Y_a +
Y_a^\dag \left[ 2 Y_a  m_{\widetilde Q_R}^2 + 4 Y_b m_{\phi a b}^2 + 4 m_{\widetilde
Q_L}^2 Y_a \right] \nonumber \\
&& + 4 A_{a}^\dag A_{a} - \frac{1}{3} M_V^2 g_V^2 - 6 M_R^2 g_R^2 -
\frac{32}{3} M_3^2 g_3^2 - \frac{1}{8} g_V^2 S_3 \nonumber \\
16 \pi^2 \frac{d}{d t} A_a &=& A_a \left[ 2 Y_b^\dag Y_b - \frac{1}{6} g_1^2 - 3 g_L^2 - 3 g_R^2 - \frac{16}{3} g_3^2 \right] + 2 Y_b Y_b^\dag A_a + 4 A_b Y_b^\dag Y_a \nonumber \\
& &+\ Y_a \left[ 4 Y_b^\dag A_b + \frac{1}{3} g_1^2 M_1 + 6 g_L^2 M_L + 6 g_R^2 M_R + \frac{32}{3} g_3^2 M_3^2 \right] \nonumber \\
& &+\ 3 {\rm Tr} \left( Y_a Y_b^\dag \right) A_b + 6 {\rm Tr} \left( A_a Y_b^\dag \right) Y_b \nonumber \\
S_3 &\equiv& 4 {~\rm Tr}\left( m_{\widetilde Q_L}^2 - m_{\widetilde
Q_R}^2 - m_{\widetilde L_L}^2 + m_{\widetilde L_R}^2 \right) + 12
\left( m_{\Delta_L}^2 - m_{\delta_L}^2 - m_{\Delta_R}^2 +
m_{\delta_R}^2 \right)
\end{eqnarray}
where $a,b=u,d$, $M_3$ is the gluino soft mass and $g_3$ is the corresponding coupling,
and $t = \ln \nu$ ($\nu$ is the running scale). The trace in $S_3$ runs over flavor
space. We first run down the energy scale in the weak basis where the up-type quark masses are diagonal $h_u \sim
M_u = {~\rm diag}\{m_u, m_c, m_t \}$, while $h_d \sim M_d = V_{\rm CKM} {~\rm diag}\{m_d,
m_s, m_b \}\ V_{\rm CKM}^\dag$. Then we turn to the super-KM basis by making the
rotation to the down-type squark masses $M^{2}_{d\ AB} \rightarrow V_{\rm CKM}^{\dag}
M^{2}_{d\ AB} V_{\rm CKM},\ A,B=L,R$. In this basis, we align the squark flavors with the
corresponding quark mass eigenstates, so all the neutral current quark-squark-neutralino
vertices are flavor diagonal, while all the charged current quark-squark-wino vertices
are proportional to the CKM mixing. The simple one-loop iteration of the RGE yields
explicit squark mass matrices in Eq. (\ref{A2}).

We study the new contributions to the kaon mass splitting and $\epsilon_K$ parameter numerically.
The free parameters are $(v_R,~m_0,~m_{1/2},~A_0,~\tan\beta,~\mu)$. $M_*$ is assumed to be around the GUT scale $10^{16}$ GeV. If demanding the new contributions not to exceed the experimental values, the date on $\epsilon_K$ constrains the argument of the effective $\Delta S=2$ matrix element to $\cal{O}$$(10^{-3})$. This requires a cancelation among contributions from $\mathscr H_{12}^{LR}$ of the $W_L-W_R$ box-diagram and from $\mathscr H_{12}^{\widetilde g}$. Numerically, the allowed parameter spaces of ($M_{W_R}$, $m_0$), ($M_{W_R}$, $m_{1/2}$), ($m_{1/2}, m_0)$, and ($M_{W_R}$, $A_0$) for $\tan\beta = 10$ and 40 are shown in Fig.~1. It is quite clear that there are large parameter spaces to lower the right-handed scale $M_{W_R}$ down to 2 TeV scale range, largely independent of the detailed values of $m_{1/2}$ and $A_0$.  For $M_{W_R}$ scale as low as several TeV, there is an upper bound on $m_0 < 600$ GeV for $\tan\beta=10$ and $m_0<800$ GeV for $\tan\beta=40$. Physically, this cancelation happens because the sign of the LR gauge-boson box diagram can be chosen independently due to the $V_R^{\rm CKM}$ in Eq.~(\ref{10}). 

\begin{figure}
\begin{center}
\includegraphics[width=5.1cm]{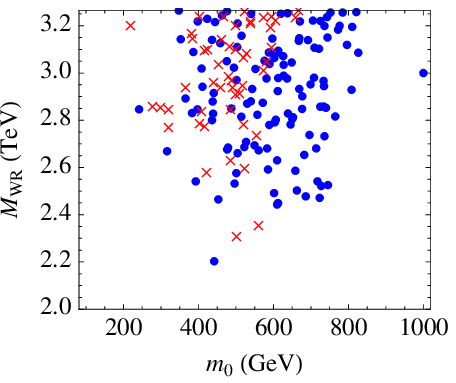}
\includegraphics[width=4.9cm]{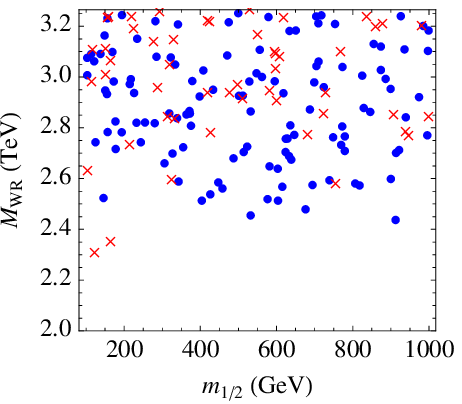} \\
\includegraphics[width=5.1cm]{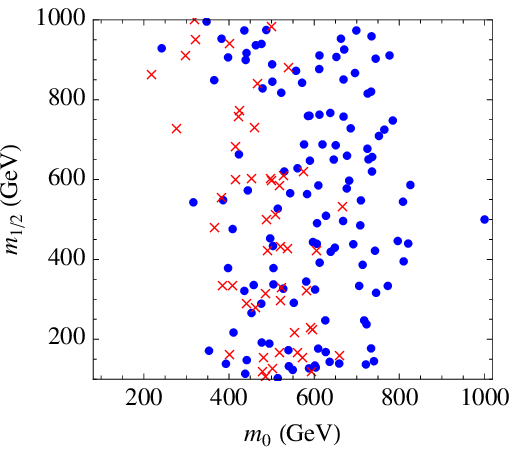}
\includegraphics[width=4.9cm]{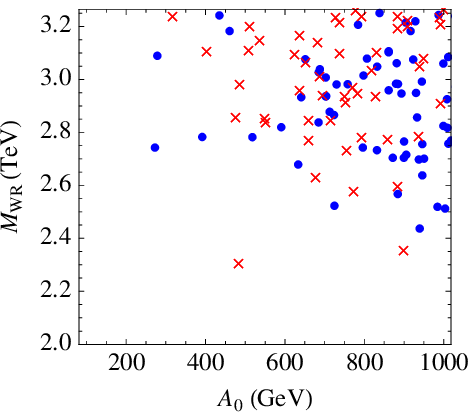}
\caption{Scatter plots of allowed parameter spaces with constraint from $\Delta m_K$ and
$\epsilon_K$. The blue dots corresponds to $\tan\beta=40$ and the red crosses $\tan\beta=10$. The correlations are shown in ($M_{W_R}$, $m_0$), ($M_{W_R}$, $m_{1/2}$), ($m_{1/2}, m_0$) and ($M_{W_R}$, $A_0$) planes.}
\end{center}
\end{figure}

\begin{figure}
\begin{center}
\includegraphics[width=5cm]{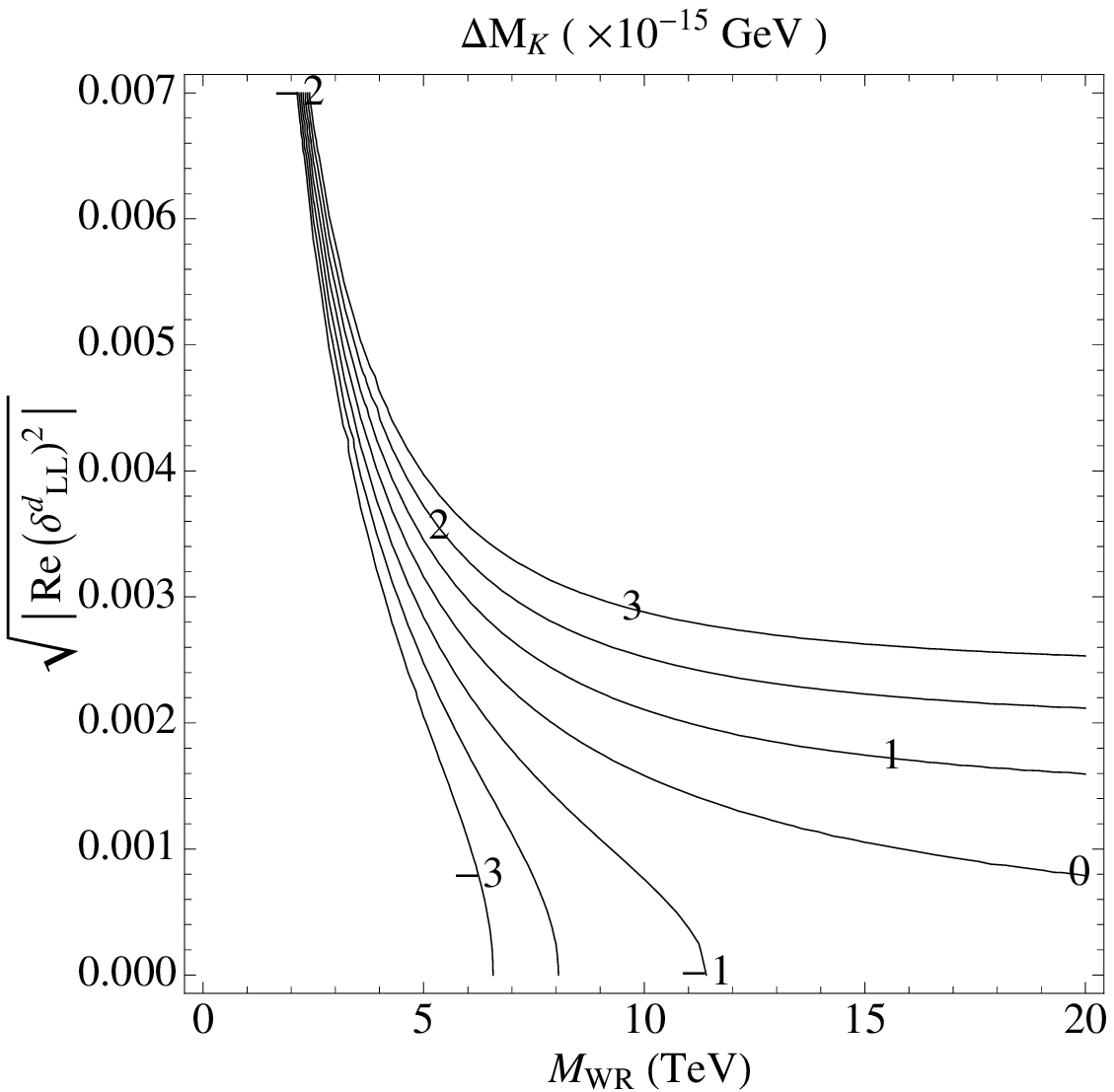}
\includegraphics[width=5cm]{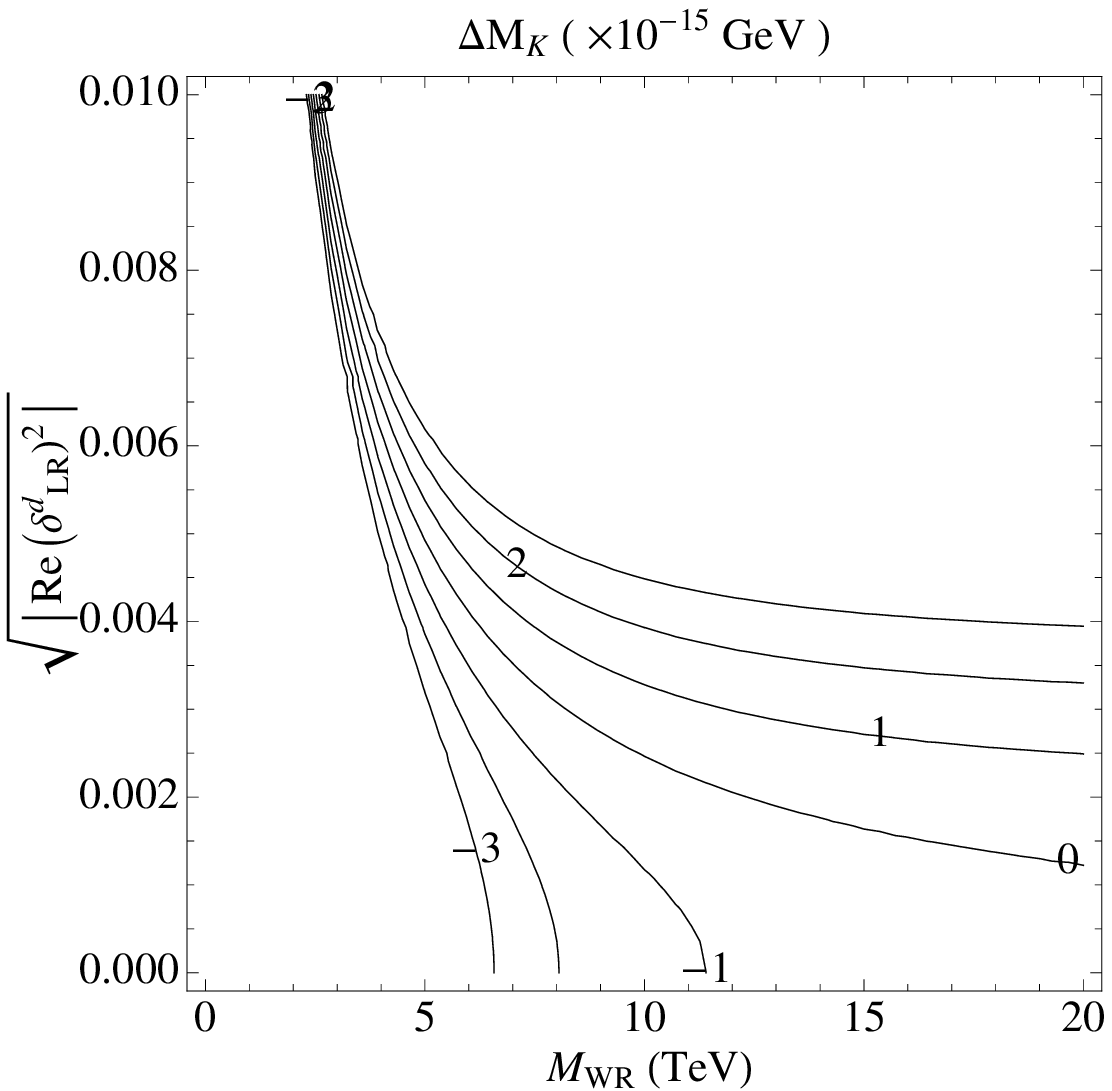} \\
\includegraphics[width=5cm]{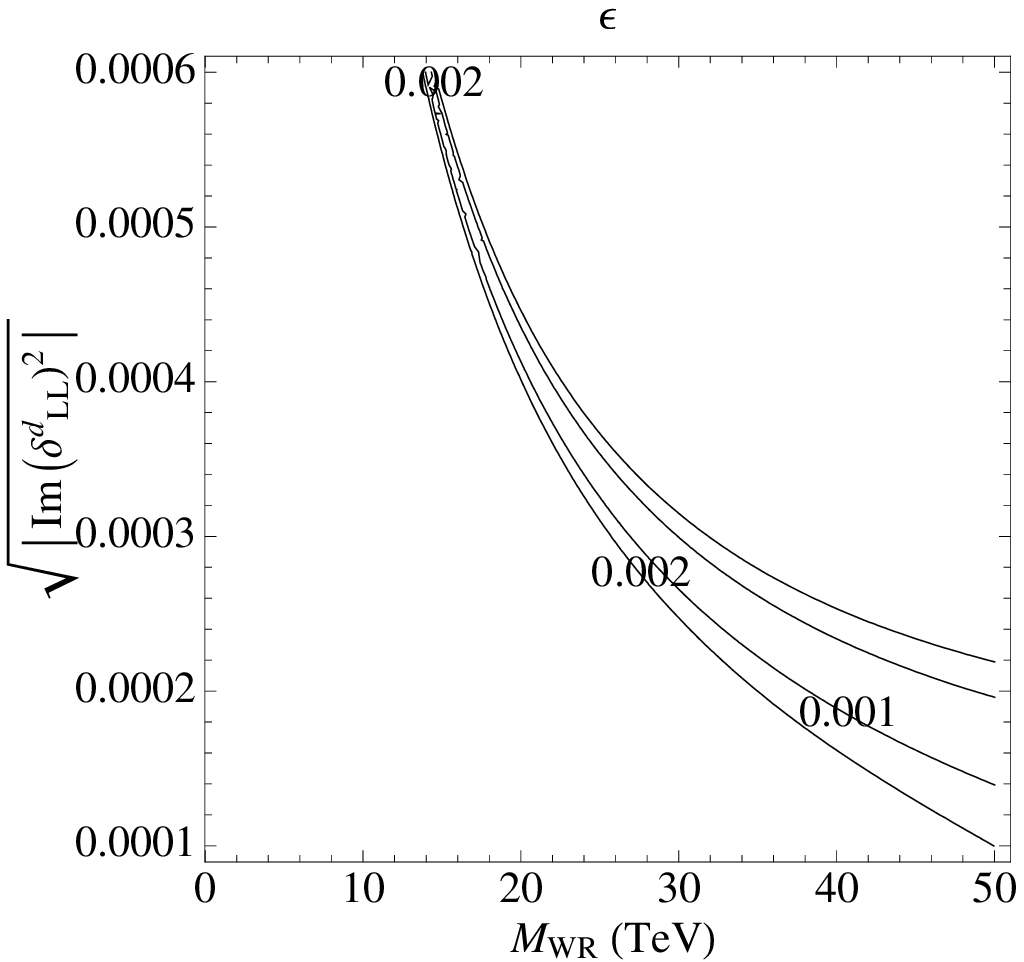}
\includegraphics[width=5cm]{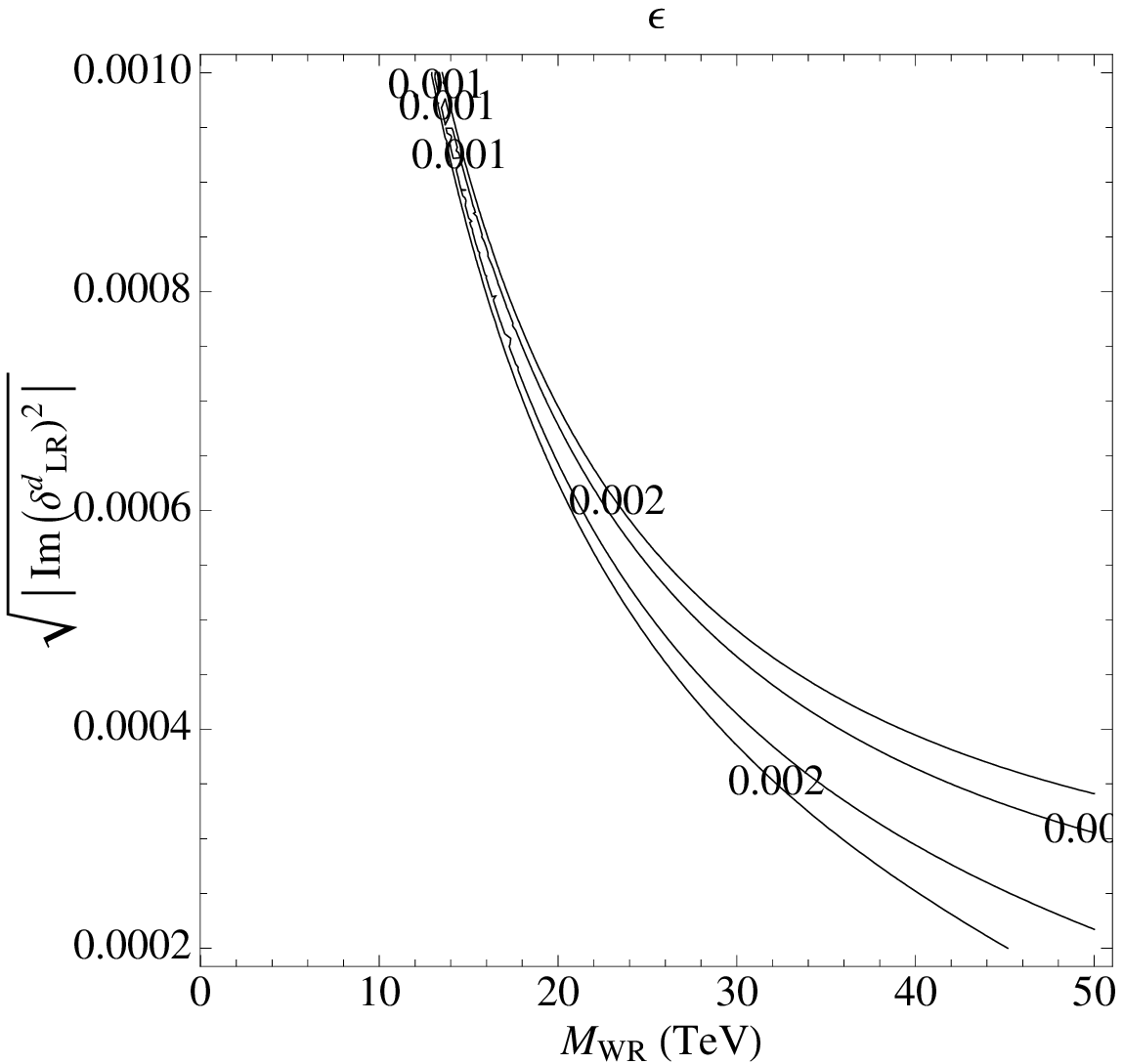}
\caption{Contour plots from constant $\Delta M_K$ and $\epsilon_K$ with FCNH contributions included. The experimental constraints can be found as bands which are functions of $M_{W_R}$ and squark mass-insertions. There is no real limit on $M_{W_R}$ if the mass insertion parameter is allowed to vary. }
\end{center}
\end{figure}

In the above discussion, we have assumed the tree-level FCNC effects are small by letting the relevant Higgs boson
masses be very large. This doublet-doublet splitting can be done by introducing a singlet or non-renormaizable terms~\cite{Mohapatra:2008gz}. This might be difficult to achieve if one demands the corresponding couplings less  than 1. The most likely possibility is that the neutral Higgs boson masses are also on the order of right-handed symmetry breaking scale. If the contribution from $v_R$ scale FCNH is included, it is difficult to get similar cancelation within the framework of mSUGRA conjecture. However, we can consider the suqark mass-insertions 
in more generic models of SUSY breaking. Including the contributions from the tree-level FCNH, $W_L-W_R$ and squark box-diagrams, cancelation among different contributions can always be achieved through adjusting off-diagonal elements of $\delta$'s without the universal boundary condition, as shown in Fig. 2. In the plots, the FCNH masses are set to equal to $W_R$ and $m_0= m_{1/2}=500$ GeV. For large $v_R$ scale, one generally gets 
upper-bounds on real and imaginary parts of $\delta$'s as in MSSM. For lower $v_R$, the magnitudes 
of $\delta$'s are constrained to be in narrow ranges.

So far in our discussion, we have neglected the chargino contribution by assuming it is small due to small
gauge and Yukawa couplings. However, in Ref.~\cite{Khalil:2001wr}, it was shown that the contribution can be
large if the charginos are light, and then $\epsilon_K$ and $\epsilon'$ can be explained consistently
with various flavor mixing parameters. In the minimal SUSY LRSM, there are
5 Dirac charginos and 9 Majorana neutralinos. Their couplings to $d$ and $s$ quarks are indeed
suppressed by the Yukawa or weak gauge couplings. The renormalization group running between $M_*$ and $v_R$
generally makes the gluino heavier than  charginos and neutralinos, and the latter contributions can
be important in some cases. A thorough analysis
of the pure SUSY contribution to kaon mixing, including the gaugino and chargino contributions has been
carried out in Ref. \cite{Frank:2003yi}. The chargino's contribution is proportional to
$(\delta^u_{LL})_{12}^2$. The study found that the upper bound on $\sqrt{{\rm Re}[(\delta_{LL}^{u})_{12}]^2}$ is less stringent than that on $\sqrt{{\rm Re}[(\delta_{LL}^{d})_{12}]^2}$ by an order of magnitude. With
the universal boundary conditions of SUSY parameters at $M_*$ and manifest LRS symmetry $V^R_{\rm CKM} = V^L_{\rm CKM}$, $(\delta_{LL}^{u})_{12}^2$ and $(\delta_{LL}^{d})_{12}^2$ have the same order of magnitude.
Therefore, in the scenario we are considering, the SUSY gluino diagrams dominate.

In conclusion, we have studied FCNC and CP-violation observables in the minimal SUSY LR
model. With the new contribution from the SUSY gluino diagrams, the right-handed $W$-boson
mass scale can be lowered to 2 TeV range where the right-handed gauge boson can be copiously 
produced at LHC . We also found a corresponding upper bounds on
the soft mass parameter $m_0$ in the model.\\

We thank R. N. Mohapatra for many useful discussions, and O. Lebedev for pointing out
possible large contribution from the charginos. This work was supported by the U. S.
Department of Energy via grant DE-FG02-93ER-40762. Y. Zhang acknowledges a support from
NSFC grants 10421503 and 10625521.

\appendix{

\section{Squark Mass Matrix in mSUGRA}

The off-diagonal terms in $m_{\widetilde Q_L}^2$, $m_{\widetilde Q_R}^2$, $A_u$ and $A_d$
appear only at lower energy in RGEs due to CKM mixng. We will neglect the running below
the righthanded scale since we are mostly interested in low-scale left-right symmetry. We
iterate once to solve equation (9), taking $m_{\widetilde Q_L}^2 = m_{\widetilde Q_R}^2 =
m_0^2\delta_{ij}$, $m_{\phi_{ab}}^2 = m_0^2 \delta_{ab}$, $ M_L = M_R = M_V = M_3 =
m_{1/2}$, $A_a = A_0 h_a$ on
the righthand side of (\ref{rge}). The leading order solution reads\ ,
\begin{eqnarray}
m_{\widetilde Q_L}^2 = \bar m_0^2 - \frac{1}{8\pi^2}( 3 m_0^2 + A_0^2 )( h_u h_u^{\dag} +
h_d h_d^{\dag} ) \ln \left(
\frac{M_{*}^2}{v_R^2} \right) \nonumber \\
m_{\widetilde Q_R}^2 = \bar m_0^2 - \frac{1}{8\pi^2}( 3 m_0^2 + A_0^2 )( h_u^{\dag} h_u +
h_d^{\dag} h_d ) \ln \left( \frac{M_{*}^2}{v_R^2} \right) \nonumber
\end{eqnarray}
where $\bar m_0^2$ contains additional flavor diagonal contribution from RG running. At
low energy, the mass spectrum for squarks can be written as
\begin{eqnarray}
- \mathcal {L}_{\widetilde q-mass} &\sim&  ( \widetilde{u}_L^{\dag},
\widetilde{u}_R^{\dag} ) \left(\begin{array}{cc}
M^{2}_{uLL}, & M^{2}_{uLR} \\
M^{2}_{uRL}, & M^{2}_{uRR} \\
\end{array} \right)
\left(\begin{array}{cc}
\widetilde{u}_L \\
\widetilde{u}_R \\
\end{array} \right) \nonumber \\
&+& ( \widetilde{d}_L^{\dag}, \widetilde{d}_R^{\dag} )
\left(\begin{array}{cc}
M^{2}_{dLL}, & M^{2}_{dLR} \\
M^{2}_{dRL}, & M^{2}_{dRR} \\
\end{array} \right)
\left(\begin{array}{cc}
\widetilde{d}_L \\
\widetilde{d}_R \\
\end{array} \right)
\end{eqnarray}
where $M^{2}_{q LR} = M^{2 \dag}_{q RL}$ and we can define $\left(
M^{2}_{q AB} \right)_{ij} \equiv m_{\widetilde q}^2
(\delta^q_{ij})_{AB}, \ A,B=L,R$. So
\begin{widetext}
\begin{eqnarray}\label{A2}
\left( M^{2}_{uLL} \right)_{ij} &=& \left( m_{\widetilde Q_L}^2
\right)_{ij} + \left[ m_{u_i}^2 + \left( \frac{1}{2} - \frac{2}{3} \sin^2 \theta_W \right) M_Z^2 \cos 2 \beta \right] \delta_{ij} \nonumber \\
\left( M^{2}_{uRR} \right)_{ij} &=& \left( m_{\widetilde Q_R}^2
\right)_{ij} + \left[ m_{u_i}^2 + \left( \frac{1}{2} - \frac{2}{3} \sin^2 \theta_W \right) M_Z^2 \cos 2 \beta \right] \delta_{ij} \nonumber \\
\left( M^{2}_{uLR} \right)_{ij} &=& A_u^{ij*} \kappa_u - \mu_{22} \left(V_L^{\rm CKM} \hat M_{d} V_R^{\rm CKM\dag}\right)_{ij} \nonumber \\
\left( M^{2}_{dLL} \right)_{ij} &=& \left( m_{\widetilde Q_L}^2
\right)_{ij} + \left[ m_{d_i}^2 + \left( - \frac{1}{2} + \frac{1}{3} \sin^2 \theta_W \right) M_Z^2 \cos 2 \beta \right]\delta_{ij} \nonumber \\
\left( M^{2}_{dRR} \right)_{ij} &=& \left( m_{\widetilde Q_R}^2
\right)_{ij} + \left[ m_{d_i}^2 + \left( - \frac{1}{2} + \frac{1}{3} \sin^2 \theta_W \right) M_Z^2 \cos 2 \beta \right]\delta_{ij} \nonumber \\
\left( M^{2}_{dLR} \right)_{ij} &=& A_d^{ij*} \kappa_d - \mu_{11} \left(V_L^{\rm CKM\dag} \hat M_{u} V_R^{\rm CKM}\right)_{ij}
\end{eqnarray}
\end{widetext}

}

\end{document}